# Semantic Information Measure
# with Two Types of Probability
# for Falsification and Confirmation[1]

Chenguang Lu

Survival99(a)gmail.com

Home Page: http://survivor99.com/lcg/english

**Abstract:** Logical Probability (LP) is strictly distinguished from Statistical Probability (SP). To measure semantic information or confirm hypotheses, we need to use sampling distribution (conditional SP function) to test or confirm fuzzy truth function (conditional LP function). The Semantic Information Measure (SIM) proposed is compatible with Shannon's information theory and Fisher's likelihood method. It can ensure that the less the LP of a predicate is and the larger the true value of the proposition is, the more information there is. So the SIM can be used as Popper's information criterion for falsification or test. The SIM also allows us to optimize the true-value of counterexamples or degrees of disbelief in a hypothesis to get the optimized degree of belief, i. e. Degree of Confirmation (DOC). To explain confirmation, this paper 1) provides the calculation method of the DOC of universal hypotheses; 2) discusses how to resolve Raven Paradox with new DOC and its increment; 3) derives the DOC of rapid HIV tests: DOC of "+" =1-(1-specificity)/sensitivity, which is similar to Likelihood Ratio (=sensitivity/(1-specificity)) but has the upper limit 1; 4) discusses negative DOC for excessive affirmations, wrong hypotheses, or lies; and 5) discusses the DOC of general hypotheses with GPS as example.

## 1. Introduction

Popper's method of falsification (1935/1959; 1963/2005) uses semantic information as criterion to test and evaluate hypotheses. Therefore, it needs a proper Semantic Information Measure (SIM). Modern inductive method (Hempel, 1945; Carnap, 1952; Eells, 2000; Hawthorne, 2004/2012) on the other hand, uses samples to confirm hypotheses. As a result, it needs a proper confirmation measure, i. e., Degree of Confirmation (DOC). There have been many SIMs (Bar-Hillel and Rudolf, 1952; Klir, 2005; Floridi, 2004, 2005/2015; Adriaans ,2010; D'Alfonso, 2011) and DOCs (Carnap, 1952; Popper, 1963/2005, 388; Earman,1992; Milne, 1996; Joyce, 1999; Christensen. 1999; Fitelson, 1999; Tentori et al, 2007). However, this paper tries to provide more reasonable SIM and DOC, so that the two methods (Falsification and Confirmation) are mutually compatible and moreover, compatible with Shannon's Information Theory (1948) and Fisher's Likelihood method. (Aldrich, 1997)

After Akaike (1974) revealed relationship between Fisher's LM and Kullback-Leibler divergence (1951), an information measure, some researchers, including the author, realized that we should use a number of samples instead of one to construct both SIM and likelihood for falsification and confirmation. (Hawthorne, 2004/2014)

The author had proposed a SIM with sampling distribution (conditional LP) and Fuzzy Truth Function (FTF, conditional LP) (Lu, 1991, 1993, 1999). This measure is compatible with Shannon's Information Theory and Popper's Falsification Theory or Hypothesis-testing Theory. Recently, the author found that this measure were also compatible with Fisher's LM and could be used for DOC. In researching about birds' sexual selection, the author proposed a hypothesis that the colorful plumages of many birds reflect their demands for foods. While measuring the semantic information of hypothesis "Birds with yellow feather like eating nectar or pollen", the author found that he could modify the true value of counterexamples or Degree of Disbelief (DOD) in a hypothesis to reduce information loss from the counterexamples and increase the average semantic information. The Degree of Belief (DOB) is determined by the equation DOD=1-|DOB|. The DOC is defined as the optimized DOB with SIM as criterion. Hence the author concludes that for any universal hypothesis, given a sampling distribution, the optimized DOB, i.e. DOC, exists and allows the average semantic information to reaches its upper limit: Kullback-Leibler Information, a special case of Shannon Mutual Information.

The methods introduced in this paper are largely different from those popular methods. First, this paper strictly distinguishes Statistical Probability (SP, denoted by $P$) from Logical Probability (LP, denoted by $T$), but use two types of probability together to test and confirm hypotheses. The $P$ means probability in which an event occurs, and $T$ means probability in which a hypothesis is judged true by different people or in different cases. The SP is divided into objective probability (in which real events occur) and subjective probability or likelihood (predicted by hypotheses or their truth functions).

Second, this paper strictly distinguishes the LP of a predicate (the less, the better) from the true value of the corresponding proposition (the larger, the better), and avoids talking "the logical probability of a proposition".

In this paper, small letters $e_1$, $e_2$… $e_m$ denote different individuals or evidences in set $A$; capital $E$ denotes a variable taking a value from $A$; that is $E \in A = \{e_1, e_2… e_m\}$. $E = e_i$ means that $e_i$ occurs. Similarly, $H$ denotes one of hypotheses or predicates $h_1$, $h_2… h_n$ in set $B$, and $H \in B = \{h_1, h_2… h_n\}$. $H = h_j$ means that $h_j$ is selected. After $h_j$ is selected, if $E = e_i$, then there is proposition $h_j(e_i)$. $A_j$ denotes a fuzzy subset of $A$ (Zadeh, 1965) so that the membership grade $\in [0,1]$ of $e_i$ in $A_j$ is the fuzzy true value of $h_j(e_i)$, denoted by $T(h_j|e_i)$ or $T(A_j|e_i)$. Hence the truth function of predicate $h_j(E)$ is $T(h_j|E)$ or $T(A_j|E)$. To resove Bar-hillel-Carnap Paradox (BCP), Floridi (2004) emphasizes to use truthlikeness to measure semantic information. The fuzzy true-value $T(A_j|e_i)$ is the truthlikeness used by the author.

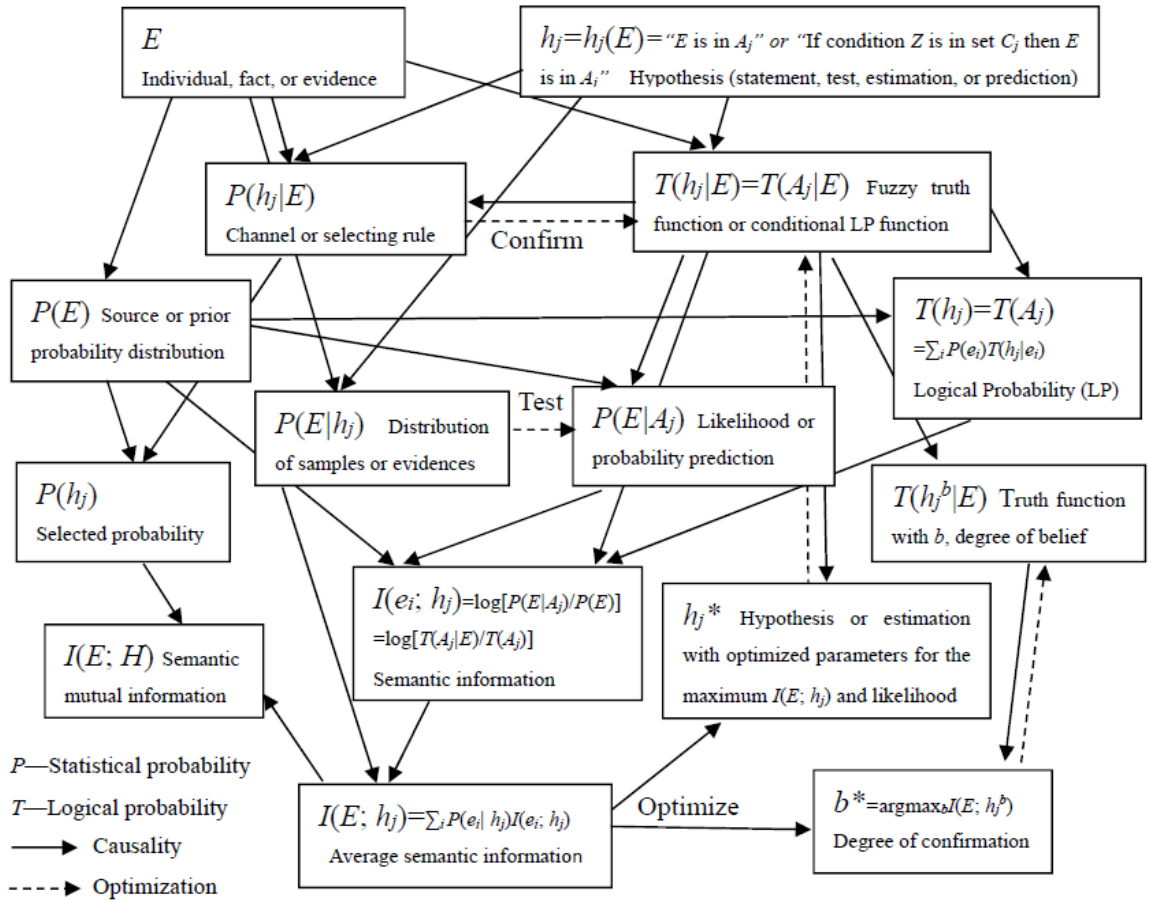

**Figure 1** Two types of probability for Information and Confirmation

——A new frame for hypothesis testing and induction[2]

Figure 1 shows how two types of probability are put together to form semantic information measure for tests, confirmations, and optimizations of hypotheses. The solid arrows means that we may get an item from other items. For example, we may get LP $T(A_j)$ from $P(E)$ and $T(A_j|E)$. Yet $P(h_j|E)$ is an exception. In some cases, $P(h_j|E)$ that already exists is prior unknown and can only be obtained by experiments, or derived from $P(E)$, $P(E|h_j)$, and $P(h_j)$ by Bayes' formula. The dotted arrows means the test, confirmation, or optimization of a hypothesis by objective conditional probability functions.

This paper first introduces the content in Figure 1, beginning from Fuzzy Truth Function (FTF); and then discusses the calculations of DOC in various cases, which is the main task of this paper.

---

## 2. Fuzzy Truth Function, Logical Probability, Statistical Probability, and Likelihood

In daily language, the truth-falsity of a statement is generally fuzzy. For example, statement "The thief is about 20 years old" is fuzzy. Its true value should be between 0 and 1. If the thief is actually 20 years old, the true value is 1. If there is deviation, true value will be less. For example, if his age is 25 years, the true value is about 0.8. If his age is 50 years, the true value is close to 0. So, the interval of true values is [0, 1] instead of {0, 1}. In the following sections, the "truth function" means FTF in most cases.

A typical example of semantic communication is weather forecast. Let $E$ denote rainfall and $H$ denote a rainfall forecast. A possible forecast is $h_j$="There will be moderate or heavy rain tomorrow". Another typical example of semantic communication is numerical prediction or estimation: "$E$ is about $e_j$" which may be written in $h_j(E)$ ="$E \approx e_j$". In mathematics, "$E \approx e_j$" is often denoted by $\hat{e}_j$. Estimations include not only those made in natural language and mathematics, but also the arrow of the Global Positioning System (GPS), the needles of a watch, or the indicators of various meters, and even a color sensation, as shown in Table 1.

Zadeh (1965) uses membership function $m_{A_j}(E)$ to define a fuzzy set $A_j$. The membership function also means the truth function of a predicate $h_j(E)$ ="$E \in A_j$". This paper follows Zadeh to define truth function $T(h_j|E)$. To emphasize the semantic meaning $h_j(E)$ = "$E \in A_j$", we also write $T(h_j|E)$ as $T(A_j|E)$ in some cases. That is to define

$$T(h_j \mid E) = T(A_j \mid E) = m_{A_j}(E) \qquad (1)$$

This function displays as a curve as shown in Figure 2. When $E=e_i$, $T(A_j|E)$ becomes the true value $T(A_j|e_i)$ of a proposition.

**Table 1** Estimations ($h_j$="$E \approx e_j$") and their true values

| Examples | Estimation $h_j$="$E \approx e_j$" | Evidence $E$, a variable | Evidence $e_i$, a constant | $T(A_j|e_i)$, true value of $h_j(e_i)$ |
|---|---|---|---|---|
| Daily language | "The thief is about 20 years old" | Real age | 18 years old | 0.9 |
| Economical prediction | "The stock index will go up about 20% this year" | Real rising percentage | 5% | 0.3 |
| balance | Reading of a balance, such as "1KG" | Real weight | 0.9KG | 0.2 |
| GPS | Arrow ↘ on a map | Real position | Right 10 meters away | 0.8 |
| Color vision | color sensation such as yellow sensation | Real color with some dominant wavelength | Color with dominant wavelength 570 nm (typical yellow). | 1 |

We could also treat $e_j$, which makes $T(A_j|e_j)$ =1, as Idea (proposed by Ancient Greek philosopher Plato) of $A_j$, and hence membership grade $m_{A_j}(e_i)$ is similarity degree or confusion probability of $e_i$ with the Idea $e_j$. So, we use not only the occurring probability of objective messages as in the classical information theory but also subjective confusion probability.

Where do truth functions come from? The truth functions of natural language come from usages. Later, the author will prove that $T(A_j|E)$ comes from selecting rule function $P(h_j|E)$. Without knowing past $P(h_j|E)$, we could still get $T(A_j|E)$ from the statistics of a random set (Wang and Sanchez, 1982). Actually, a fuzzy hypothesis (or its truth function) is similar to a predictive model (or its parameters' set) in statistics, such as in the Maximum Likelihood Method (MLE) (Aldrich, 1997). If $h_j$ is an approximately unbiased estimation, its truth function may be approximately written as

$$T(A_j|E) = \exp\ [-(E - e_j)^2/(2d^2)] \qquad (2)$$

where $d$ is standard deviation. The larger the $d$ is, the fuzzier the estimation is. Note that the maximum of truth function $T(A_j|E)$ is 1.

Unlike other popular methods for measuring semantic information, this paper strictly distinguishes LP from SP, and uses both to measure semantic information.

First, $E$ or $e_i$ is a fact or evidence. It only has SP $P(E)$, without LP $T(E)$. In practice, if one suspects that an evidence is false, one may use a hypothesis with true value between 0 and 1 to replace it.

Second, a hypothesis $h_j = h_j(E)$ has both SP (selected probability $P(h_j)$) in which $h_j$ is selected, and also LP (denoted by $T(h_j) = T(A_j)$) in which $h_j(E)$ is judged true. They are generally different. Consider hypotheses $h_1$="There will be small rain", $h_2$="There will be moderate rain", and $h_3$="There will be small to moderate rain". According to their semantic meanings, $T(h_3) \approx T(h_1) + T(h_2)$; yet, there may be $P(h_3) < P(h_1)$. The LP of tautology is 1; yet its selected probability is close to 0.

Third, SP is normalized (an exception will be talked late), for examples, $P(e_1) + P(e_2) + ... + P(e_m) = 1$; $P(h_1) + P(h_2) + ... + P(h_n) = 1$, and $P(e_1|h_j) + P(e_2|h_j) + ... + P(e_m|h_j) = 1$; Yet, LP is not normalized and has the maximum 1. Generally, $T(A_j|e_1) + T(A_j|e_2) + ... + T(A_j|e_m) > 1$; $T(A_1) + T(A_2) + ... + T(A_n) > 1$ because $A_1$, $A_1$, ..., $A_n$ are not disjoint. Only when they are disjoint and hypotheses $h_1, h_2... h_n$ are always correctly selected, $T(A_j) = P(h_j)$, $j=1, 2... n$.

Averaging truth function, we get the LP of predicate $h_j(E)$:

$$T(A_j) = \sum_i P(e_i) T(A_j\,|\,e_i) \qquad (3)$$

This is just the fuzzy set probability defined by Zadeh. (1986)

Note that the LP of a predicate in this paper is different from the true value of a predicate in mathematical logic, which is equal to $T(h_j|e_1) \wedge T(h_j|e_2) \wedge ... \wedge T(h_j|e_m)$, because 1) the true value in Mathematical Logic can only be 0 or 1, yet the LP can be any value between 0 and 1; 2) the true value is irrelative to $P(E)$, yet the LP is related to $P(E)$; 3) the true value is posterior, yet, the LP is prior.

The LP defined by Eq. (3) is also different from the LP defined by Bar-hillel and Carnap (1952) and others (Floridi, 2004), which is also irrelative to $P(E)$. For example, according to their definition, three predicates divide the logical space into $2^3=8$ lattices and hence the minimum of the LP is 1/8. However, according Eq. (3), the LP $T(A_j)$ depends not only on the coverage of $A_j$, but also on the probability distribution of those $E$ over $A_j$. For example, although "That man is over 100 years old" has larger coverage than "That man is about 60 years old", its LP is much less because $P$(men's age>100) is very small. Therefore, the smaller LP of a hypothesis means that the event described is more specific and more occasional. It is specificity and occasionality that Popper uses to explain the severity of tests and his information criterion.

Strictly speaking, the term "logical probability of a proposition" is improper, because a proposition only has true vale rather than LP. We could treat the $T(A_j)$ as the prior LP of hypothesis $h_j$, and the true value $T(A_j|e_i)$ as the posterior LP. Because of improper usage of this term, researchers are often puzzled by the question: whether larger LP or less LP of a proposition is better? The author believes that less LP and larger true value are better (or less prior LP and larger posterior LP are better), because the less the LP is, the severer the test is; the larger the true value is, the better the hypothesis survives the test.

When $h_j$ is selected, the probability of $E$ is $P(E|h_j)$; while $h_j$ is true, probability of $E$ should be (Author, 19)

$$P(E \mid A_j) = \frac{P(E)T(A_j \mid E)}{T(A_j)} \quad (4)$$

This formula may be called semantic Bayes' formula, which establishes the relationship between SP and LP. In terms of MLE, $A_j$ (or $T(A_j|E)$) is a predictive model, $P(E|A_j)$ is the likelihood function. Note that the peak of likelihood $P(E|A_j)$ is between the peak of $T(A_j|E)$ and the peak of $P(E)$ as illustrated by Figure 2.

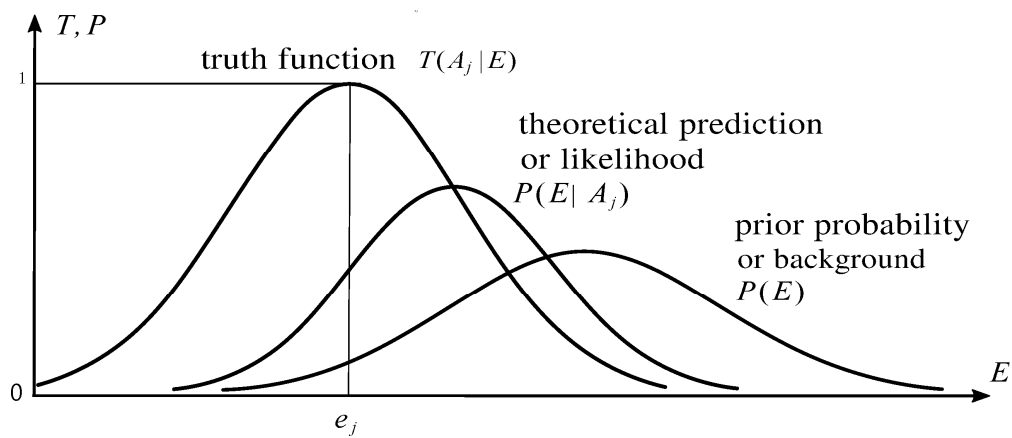

**Figure 2** Likelihood $P(E|A_j)$ locates between truth function $T(A_j|E)$ and source $P(E)$

In the popular methods, $P(E|b)$ is used as the probability distribution of $E$ for given background knowledge $b$. In this paper, equally probable distribution $P(E) \equiv 1/m$

means no background knowledge; $P(E)$ has carried background knowledge already. In other words, $P(E)$ means $P(E|b)$, or say, $b$ is omitted in this paper. The author uses $P(E)$ in a way similar to the way Shannon uses $P(X)$. (1948)

## 3. Semantic Information Measure for Falsification

According to classical information theory, relative information formula (Rosie, 1966) is:

$$I(e_i; h_j) = \log \frac{P(e_i \mid h_j)}{P(e_i)} \qquad (5)$$

This formula is the core of Shannon's mutual information formula (1948). Averaging $I(e_i; h_j)$, we will get Shannon's mutual information $I(E; H)$. However, Shannon never used this formula. The reason is that the use of this formula may bring negative information; yet Shannon's formulas of entropy and mutual information only measure mean information, which is always positive. The author believes that negative information is possible and meaningful to semantic information because if we believe lies or wrong predictions, the information will be negative.

To replace $h_j$ in Eq. (5) with $h_j$ *is true* so that Eq. (5) becomes (Author, 19    ):

$$I(e_i; h_j) = \log \frac{P(e_i \mid h_j \text{ is true})}{P(e_i)} \log \frac{P(e_i \mid A_j)}{P(e_i)} \qquad (6)$$

This formula is similar to the formula proposed by Popper for severity of tests (1963/2005, 526) and the formula proposed by Milne (1996) for DOC. However, differences are that 1) The $h_j$ is replaced by $A_j$ which clearly means that $h_j$ *is true* rather than $h_j$ *is selected*; 2) Background knowledge $b$ is omitted here. According to Eq. (4) and (6), we get the Semantic Information Formula (SIF):

$$I(e_i; h_j) = \log \frac{T(A_j \mid e_i)}{T(A_j)} \qquad (7)$$

which is illustrated in Figure 3.

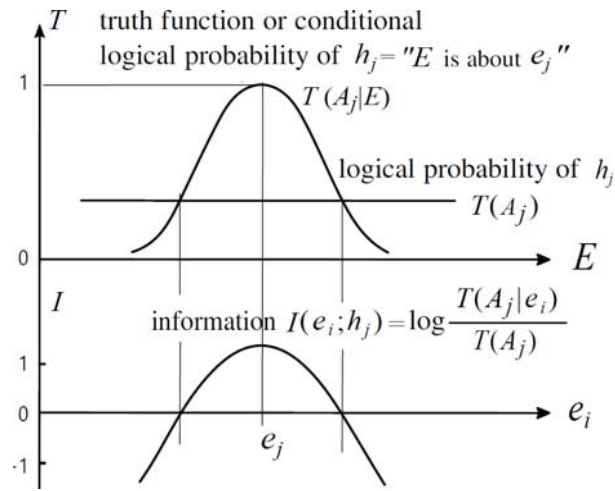

**Figure 3** The illustration of semantic information formula

Floridi (2004) follows Popper (1963/2005, 526, 534) to emphasize truthlikeness or verisimilitude for semantic information. It is $T(A_j|E)$ that is used for truthlikeness.

The above semantic information measure has three characteristics:

1. To determine the amount of information, we need to test the prediction by the evidence. When the evidence is exactly consistent with the prediction, that is $e_i=e_j$, the information reaches the maximum. The information decreases as deviation increases. When deviation reaches a certain level, information is negative. This relationship just right manifests ordinary error criterion.

2. The smaller the LP $T(A_j)$ (i. e., the lower the horizontal line in Figure 3) is and the larger the true value $T(A_j|e_i)$ is, the larger the $I(e_i; h_j)$ is. This exactly manifests Popper's notion: the smaller the logical probability of a hypothesis is, the more information there is if it can survive tests.

3. The information is 0 for a tautology or a contradiction. Popper affirms that a tautology contains no information, because it is not testable or logically non-falsifiable. The above formula reaches the same conclusion. For a tautology, $T(A_j|E)\equiv 1$, $T(A_j)=P(e_1)+P(e_2)+\ldots+P(e_m)=1$; so, $I(e_i; h_j)=\log(1/1)=0$. It is also easy to avoid Bar-hillel-Carnap Paradox (Floridi, 2004, 2005/2015) by this formula. For a contradiction, $T(A_j|E)\equiv 0$ and $T(A_j)=0$. So, $I(e_i; h_j)=\log(0/0)$. Since $\log(0_+/0_+)$ $=\log1=0$ ($0_+$ is an infinitesimal). So, it is reasonable to think $I(e_i; h_j)=0$ for a contradiction.

Now let's consider the information of estimation "$E \approx e_j$" with the truth function $T(A_j|E)=\exp[-(E-e_j)^2/(2d^2)]$. The SIF may be written as

$$I(e_i; h_j) = \log[T(A_j | e_i)/T(A_j)] = \log[1/T(A_j)] - (e_i - e_j)^2/(2d^2) \quad (8)$$

The above formula may be understood as *Information= Testing severity - Relative deviation.*

Popper defined Testing severity and Verisimilitude (1963/2005, 526, 534). Since LP and SP are not well distinguished by him, his definitions are not satisfactory. The author suggests defining log $[1/T(A_j)]$ as testing severity, and $T(A_j|e_i)/T(A_j)$ as verisimilitude. In terms of LM, $P(e_i|A_i)/P(e_i)=T(A_j|e_i)/T(A_j)$ is also called standard likelihood. So, we may say

Semantic information = log (Standard likelihood)

= log (Verisimilitude)=Testing severity - Relative deviation

If negative verisimilitude for lies or wrong predictions is expected, one may also define verisimilitude by log $[T(A_j|e_i)/T(A_j)]$.

Averaging $I(e_i; h_j)$ for different $i$ in Eq. (7), we can obtain Average Semantic Information (ASI) of hypothesis $h_j$:

$$I(E; h_j) = \sum_i P(e_i | h_j) \log \frac{T(A_j | e_i)}{T(A_j)} \quad (9)$$

which is called Average Semantic Information Formula (ASIF) where $P(e_i| h_j)$, $i$=1, 2… $m$ is sampling distribution from statistics as a group of evidences. According to this formula, if there is a counterexample $e_i$ for which $P(e_i| h_j)$>0 and $T(A_j|e_i)$ =0, yet $T(A_j)$>0, then the average information is -∞. This coincides with Popper's assertion: One exception is enough to falsify a universal hypothesis. However, this assertion can only be applied to non-fuzzy hypotheses. How can we test fuzzy hypotheses (such as "People with high Triglyceride probably also have fatty liver" and "There will be small to moderate rain tomorrow")? Popper did not offer a proper method. In daily life and the field of Social Science, most hypotheses are fuzzy. The above formula allows a reasonable evaluation of these hypotheses under the frame of Popper's theory and avoids negative infinite information.

The above formula may also be written as

$$I(E; h_j) = \sum_i P(e_i \mid h_j) \log \frac{P(e_i|A_j)}{P(e_i)} \qquad (10)$$

where $P(e_i| A_j)$, $i$=1, 2… $m$ is likelihood function and may be understood as theoretical prediction; $P(e_i)$, i. e., $P(e_i|b)$, $i$=1, 2… $m$ may be understood as prior likelihood, background knowledge, or context. This formula is a generalization of Kullback-Leibler (KL) formula (1951) and may be called Generalized Kullback-Leibler Formula (GKLF), which is illustrated in Figure 4. The information measured by Eq. (9) and (10) also has coding meaning (Lu, 1994, 2012).

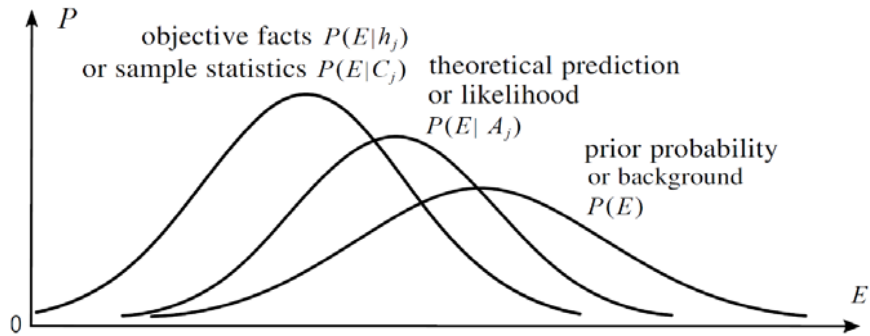

**Figure 4** The illustration of generalized Kullback-Leibler formula

The $I(E; h_j)$ may be called generalized KL information. It can be written as the difference of two KL divergences:

$$I(E; h_j) = \sum_i P(e_i \mid h_j) \log \frac{P(e_i|h_j)}{P(e_i)} - \sum_i P(e_i \mid h_j) \log \frac{P(e_i|h_j)}{P(e_i \mid A_j)} \qquad (11)$$

Since KL divergence is larger than or equal to 0, the information reaches its maximum when

$$P(e_i \mid A_j) = P(e_i \mid h_j) , i=1, 2… \ m \qquad (12)$$

so that the second part is 0. The maximum is equivalent to the KL information. Therefore, KL information is the upper limit of generalized KL information. First, the generalized KL information conforms to ordinary error criterion: Consistency is good. Second, it ensures that the more different $P(E|h_j)$ is from $P(E)$, the more information is conveyed if $P(E|A_j)$ is close to $P(E|h_j)$. This is the very manifestation of Popper's viewpoint: the more unexpected a prediction is and thus the severer tests it undergoes, the more information it conveys if it can survive tests.

## 4. Compatibility between Generalized KL Information and Maximum Likelihood Estimation

Akeike (1974) revealed the relationship between KL formula and MLE. The author will explain the relationship between GKLF and MLE below.

Let $Z$ denote an observed condition, $C=\{z_1, z_2 \ldots z_w\}$ be a set of independent conditions, and $Z \in C$. For given $Z=z_k$, the conditional probability of $E$ is $P(E|z_k)$. Assume some elements in $C$ result in the similar $P(E|.)$, we merge these elements into a subset $C_j$ of $C$. Then if, and only if $Z \in C_j$, we select $h_j$. Hence $P(E|h_j) = P(E|Z \in C_j)$, denoted by $P(E|C_j)$, so that the GKLF becomes

$$I(E; h_j) = \sum_i P(e_i | C_j) \log \frac{P(e_i | A_j)}{P(e_i)} \qquad (13)$$

The likelihood (function) of a predictive model $\theta$ (or its parameters' set) is defined as $L(\theta|E) = P(E|\theta)$. To train the model, we use $w$ samples $e(1), e(2) \ldots e(w) \in A$. Then the likelihood becomes $L(\theta|E^w) = P(E^w|\theta)$. If these samples are independent, then

$$P(E^w|\theta) = P(e(1)|\theta)P(e(1)|\theta) \ldots P(e(w)|\theta) \qquad (14)$$

Assume there are $w_i$ samples that are $e_i$ among $w$ samples, $i=1, 2 \ldots m$, hence

$$P(E^w | \theta) = \prod_i P(e_i | \theta)^{w_i} \qquad (15)$$

Assume these samples occur under condition $Z \in C_j$, and $w$ is enough large. Hence $P(e_i|C_j) = w_i/w$; the logarithm of $P(E^w|\theta)$ becomes

$$\log P(E^w | \theta) = w \sum_i P(e_i | C_j) \log P(e_i | \theta) \qquad (16)$$

Compare Eq. (13) and (16), it is easy to find that a fuzzy set $A_j$ is equivalent to a model $\theta$. Seeking a truth function $T(A_j|E)$ with optimal parameters that result in the maximum $I(E; h_j)$ is equivalent to seeking the optimal parameters of the model $\theta$ that result in the maximum likelihood. The difference is that the semantic information method separates model (truth function $T(A_j|E)$) and source $P(E)$, and gets likelihood $P(E|A_j)$ by semantic Bayes' formula Eq. (4). Yet, the likelihood method does not separate them and directly constructs the likelihood $P(E|\theta)$. So, the semantic information method can be used for MLE when source $P(E)$ is variable.

From Eq. (12), we know that the average information $I(E; h_j)$ reaches its maximum when

$$T(A_j|E) = T(A_j)P(h_j|E)/P(h_j) \qquad (17)$$

This is the inverse formula of semantic Bayes' formula (4). Assume when $E=e_j^*$, $P(h_j|E)$ has the maximum $P(h_j|e_j^*)$. Let the maximum of $T(A_j|E)$ be $T(A_j|e_j^*) = 1$, we get optimized truth function

$$T(A_j|E) = P(h_j|E)/P(h_j| e_j^*) = P(E|h_j)/P(E)/[P(e_j^*)/P(e_j^*|h_j)] \qquad (18)$$

For MLE, if the number ($w$) of samples is big enough, then the above equation becomes

$$T(A_j|E) = P(C_j|E)/P(C_j|e_j^*) = P(E|C_j)/P(E)/ [P(e_j^*)/P(e_j^*|C_j)], \quad j=1, 2\ldots n \qquad (19)$$

The above estimation method may be called the Maximum Semantic Information Estimation (MSIE). The Eq. (18) or (19) may be called Fuzzy Information Criterion (FIC) of estimations.

Note that the conditional probability function $P(H|e_j)$ is normalized; yet $P(h_j|E)$ is not normalized because the left $h_j$ is not a variable. That means that generally, $P(h_j| e_1) + P(h_j| e_2) + \ldots + P(h_j| e_m) \neq 1$ (which may be seen in Table 7). We may call $P(h_j|E)$ Selecting Rule Function of $h_j$. Note that all $P(h_j|E)$, $j=1, 2\ldots n$, form Shannon's channel $P(H|E)$. So, Eq. (18) indicates how semantic channel matches Shannon channel to convey most information.

Philosopher Weitgenstein has a famous standpoint (1958, 80): the meaning of a word lies in its use. Obviously, Eq. (18) supports this standpoint. When the audience or listeners continue to improve their understanding, forecasters or speakers also continue to improve their selecting rules of sentences, including selecting $h_j$ according to observed condition $Z$. Language is evolving in this way.

## 5. The Optimization of Degree of Disbelief

The Degree of Belief (DOB) is a degree to which one believes a hypothesis. It is subjective and prior. After the hypothesis is tested by a series of samples, one gets the optimized DOB, i.e., DOC. So, DOC is the degree of inductive support (Hawthorne, 2005). Since now, let's use $b$ to denote the DOB, $b^*$ to denote the optimized $b$ or DOC, $b'=1-|b|$ to denote the DOD, and $b'^*$ to denote the optimized DOD or degree of disconfirmation.

In his studies of aesthetics and birds' sexual selection, the author proposed a hypothesis that colorful plumages of most male birds were selected by female birds and female tastes for beauty came from and indicated their demands for foods or environment. For example, the male peacock mimics the berry tree to attract the female because they like eating berries. First, the demanding relationship between the peacock and berries selected the peacock's taste for beauty; lately, the female taste for beauty selected the male feathers. When the author tried to use statistical data and semantic information measure to test and evaluate the hypothesis $h_1=$"Birds with yellow feathers like eating nectar or pollen", he found that increasing the true value of counterexamples properly could reduce the information loss and increase the average semantic information $I(E; h_1)$. The true value of the counterexamples may be regarded as $b$', the DOD to $h_1$. The following is the introduction of this method.

In natural language, to reduce the information loss from counterexamples, we may use two methods to increase the fuzziness of hypotheses predictions. One is to use words such as "about" or "similar" to increase the fuzziness of a hypothesis by decreasing its precision. The $d$ in Eq. (8) indicates the precision. The larger the $d$ is, the lower the precision is, and hence the fuzzier the hypothesis is. Another method is to use words such as "probably" or "plausibly" to decrease audience's DOB $b$. The less the $b$ is, the fuzzier (or more like a tautology) the hypothesis is. The audience may adjust again the DOB in a hypothesis. For example, generally, people give lower DOB to economists' predictions; slightly higher DOB to weather forecasts and medical diagnoses; higher DOB to GPS, watches, and thermometers. For this reason, the DOB $b$ of a predicate $h_j(E)$ (instead of a proposition) is defined by

$$T(h_j{}^b|E) = b' + bT(A_j|E), \text{ for } b > 0 \qquad (20)$$

where $h_j$ is the initial hypothesis, which is fuzzy or non-fuzzy; $h_j{}^b$ is $h_j$ with DOB $b$; $b'=1-|b|$ is the degree of disbelief. This definition actually treats the DOD $b$' as the proportion of the tautology in the $h_j$ with $b$.

Figure 5 shows an example of modifying the true-value of counterexamples from 0 to $b$' for the non-fuzzy hypothesis $h_j$="There wil be small rain (rainfall betwee 0 and 5 mm) tomorrow".

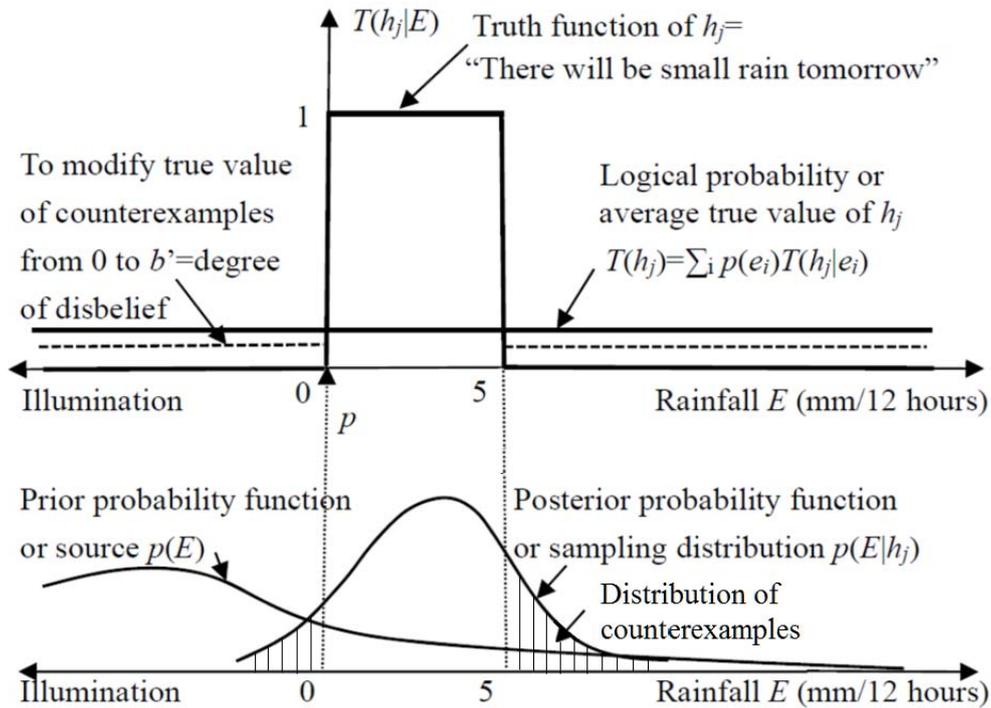

**Figure 5** Modifying the true-value of non-fuzzy hypothesis from 0 to $b$' when counterexamples exist.

A non-fuzzy universal hypothesis is defined as:

$h_1$="For all $E$, if $E$ is in $S_1$, then $E$ is also in $S_2$"

where $S_1$ and $S_2$ are two (non-fuzzy) sets. If we believe $h_1$ to some degree $b$, $h_1$ becomes a fuzzy hypothesis, denoted by $h_1{}^b$.

Consider these fuzzy inferences (with more or less counterexamples): "Birds with yellow feathers probably like eating nectar or pollen", "People with high triglyceride probably have fatty livers", "HIV-positive people are very probably infected by HIV", and "Probably all swans are white". They may be called fuzzy universal hypotheses.

Assume all evidences in $A$ can be divided into two types (as shown in Table 2): $e_1$ (in $S_2$) and $e_0$ (not in $S_2$), or four kinds: $e_{11}$ (in $S_1$ and $S_2$), $e_{00}$ (not in $S_1$ and not in $S_2$), $e_{10}$ (in $S_1$ and not in $S_2$), and $e_{01}$ (not in $S_1$ and in $S_2$).

**Table 2** Evidences divided into four types for a universal hypothesis

| $E$ | $e_1 \in S_2$ | $e_0 \notin S_2$ |
|---|---|---|
| $E \in S_1$ | $e_{11}$ | $e_{10}$ |
| $E \notin S_1$ | $e_{01}$ | $e_{00}$ |

The fuzzy universal hypothesis $h_1{}^b$ is defined as:

$$h_1{}^b = \text{"For all } E \text{, if } E \text{ is in } S_1 \text{, then } E \text{ is also in } A_1\text{"}$$

where $b$ is the DOB in $h_1$ and $A_1$ is the fuzzified $S_2$; the elements of $A_1$ makes $h_1{}^b$ true. Let $S_1$' and $S_2$' be the supplementary sets of $S_1$ and $S_2$ respectively. We define $h_0{}^{b0}$ as

$$h_0{}^{b0} = \text{"For all } E \text{, if } E \text{ is in } S_1\text{', then } E \text{ is also in } A_0\text{"}$$

where $b_0$ is the DOB in $h_0$ and $A_0$ is the fuzzified $S_2$'; the elements of $A_0$ make $h_0{}^{b0}$ true. For the universal hypothesis $h_1$, $T(h_1|e_{11}) = 1$ and $T(h_1|e_{10}) = 0$. For the fuzzy universal hypothesis $h_1{}^b$, according to Eq. (20), $T(A_1|e_{11}) = 1$ and $T(A_1|e_{10}) = b$'. Similarly, $T(A_0|e_{00}) = 1$ and $T(A_0|e_{01}) = b_0$'. The four truth values of two fuzzy inferences are shown in Table 3.

**Table 3** Four truth values of two fuzzy inferences

| | $e_1 \in S_2$ | $e_0 \notin S_2$ |
|---|---|---|
| $E \in S_1$ , $h_1{}^b$ | $T(A_1|e_{11}) = 1$ | $T(A_1|e_{10}) = b' = 1 - \|b\|$ |
| $E \notin S_1$, $h_0{}^{b0}$ | $T(A_0|e_{01}) = b_0' = 1 - \|b_0\|$ | $T(A_0|e_{00}) = 1$ |

Let $P_1 = P(e_1)$, $P_0 = P(e_0)$, $Q_1 = P(e_1|h_1)$ and $Q_0 = P(e_0|h_1)$. Using Eq. (3), we get $T(A_1) = b'P_0 + P_1$. Using Eq. (9), we get

$$I(E; h_1{}^b) = Q_0 \log \frac{b'}{b'P_0 + P_1} + Q_1 \log \frac{1}{b'P_0 + P_1} \qquad (21)$$

According to the above formula, when $b=1$ or $b'=0$ which means that the listeners fully believe $h_1$, if there is a counterexample, the information will be $-\infty$. When $b=0$ or $b'=1$ which means that the listeners do not believe $h_1$ at all, the average information is 0. We can seek $b$' (for $0 \leq b' \leq 1$) that makes $I(E; h_1{}^b)$ reach its maximum（see Figure 5）.

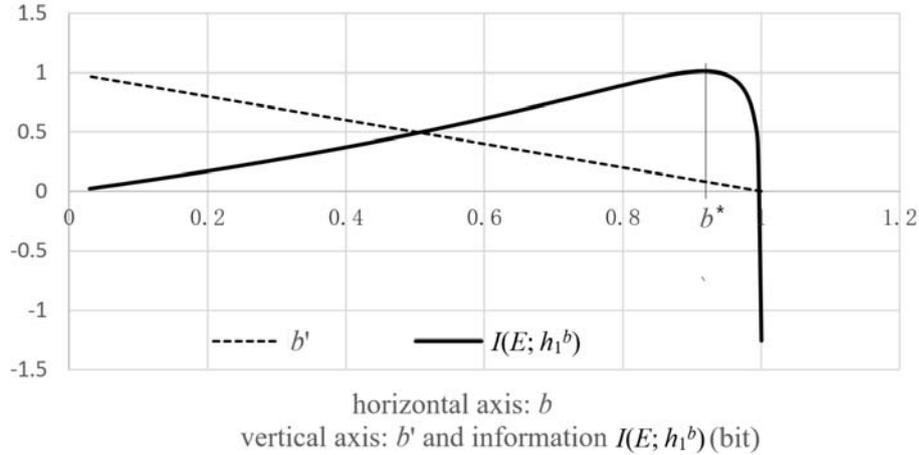

Figure 6 Information $I(E; h_1{}^b)$ changes with degree of disbelief $b$'

for $P_0/P_1$=0.8/0.2; $Q_0/Q_1$=0.25/0.75

Let derivative d$I(E; h_1{}^b)$/d$b$'=0, we get $Q_0P_1$-$Q_1P_0b$'=0. Hence there is

$$b'= (Q_0/Q_1)/(P_0/P_1) \qquad (22)$$

Since the second derivative is less than 0 for $b$'= $b$'*. So, $b$'* is the optimized $b$' that maximizes $I(E; h_1{}^b)$.

Below is the discussion about Eq. (22) only for cases where $Q_0/Q_1$<$P_0/P_1$. If $Q_0/Q_1$>$P_0/P_1$, we need another formula to get $b$'*.

When $P_0$<$P_1$, $P_0/P_1$ may be called prior Absolute Degree of Disbelief (ADOB); when $Q_0$<$Q_1$, $Q_0/Q_1$ may be called posterior ADOD. So, $b$'* may be regarded as the decrement of ADOD. If $P_0$=$P_1$=0.5, then $P_0/P_1$=1 and $b$'*= $Q_0/Q_1$ which means that without background knowledge $P(E)$, the $b$'* is equal to the posterior ADOD. From ADOD, we could get absolute conformation measure (Huber, 2005).

From Eq. (20) and (22), we get

$$b^*=1-(Q_0/Q_1)/(P_0/P_1) \qquad (23)$$

If $Q_0$=0 and $P_0$>0, then $b$'*=0 and $b$*=1, which means the hypothesis $h_1$ is completely confirmed. Eq. (22) may also be written as $b$'*=$(Q_0/P_0)/(Q_1/P_1)$ which means that DOD decreases with counterexamples' decreasing and positive examples' increasing. According to Eq. (18), we may directly get

$$b'^*= (Q_0/P_0)/(Q_1/P_1) =P(h_1|e_{10})/P(h_1|e_{11}) \qquad (24)$$

So, there is also

$$b^*=1-b'^*=1- P(h_1|e_{10})/ P(h_1|e_{11}) \qquad (25)$$

which means that the DOC of $h_1$ is only related to selecting rule function $P(h_1|E)$ and truth function $T(h_1|E)$ but source $P(E)$. Note that test or semantic information is related to $P(E)$.

The Eq. (23) looks like a formula from Likelihoodism, yet the Eq. (25) looks like a formula from Bayesianism. The Eq. (24) shows that Likelihoodism and Bayesianism (Fitelson, 2007) can be compatible when sampling distribution is used to confirm a hypothesis or its truth function (conditional LP function). However, every $P$ in above eqations means SP rather than LP. The conditional LP (function) is to be confirmed, and hence does not occur in these formulas. Otherwise, that is to let itself support itself.

With the semantic Bayes' formula (4), we can use $b'^*$ and $P(e_1)$ to calculate predicted probability of $e_1$, denoted by $P(e_1|h_1^*)$. That is

$$P(e_1|h_1^{b^*})=P(e_1)/[P(e_1)+b'^*P(e_0)] \qquad (26)$$

Figure 6 shows how $P(e_1|h_1^{b^*})$ and $I(E; h_1^{b^*})$ are positively related to $b^*$.

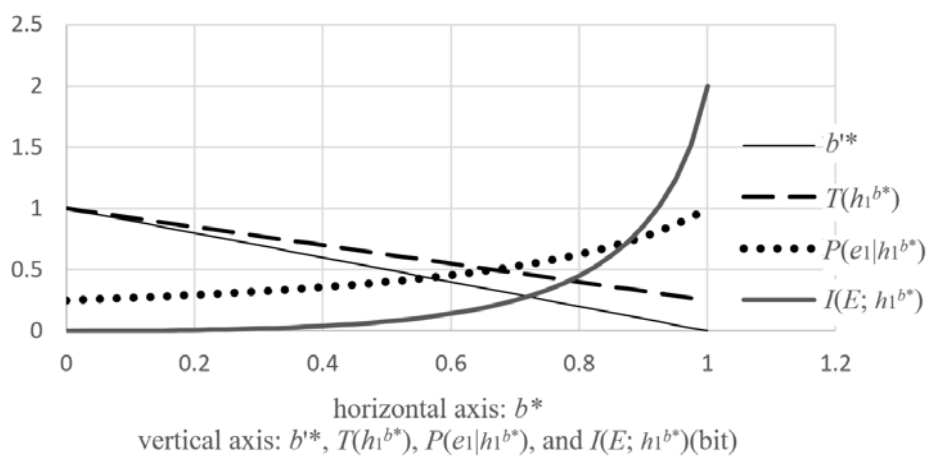

horizontal axis: $b^*$
vertical axis: $b'^*$, $T(h_1^{b^*})$, $P(e_1|h_1^{b^*})$, and $I(E; h_1^{b^*})$(bit)

**Figure 7** Relations of $b^*$ to accuracy rate $P(e_1|h_1^{b^*})$ and information $I(E; h_1^{b^*})$

for prior probability $P(e_1)$=0.2.

Now we use both $b^*$ and $I(E; h_1^{b^*})$ to evaluate and compare two hypotheses. There are 843 representative birds in the book *The Illustrated Encyclopedia of Birds of the World* (Alderton, 2005). Those birds could be divided into four types by whether they have yellow feathers (including orange feathers) and whether they eat nectar (including pollen), as shown in Table 4, where $n_{11}$ is the number of $e_{11}$, and so on.

**Table 4** The optimization of the degrees of disbelief $b'$ about birds

|  | eating nectar ($e_1$) | not eating ($e_0$) |
|---|---|---|
| with yellow feathers ($h_1$) | $n_{11}$= 83 | $n_{10}$=57 |
| without yellow feathers ($h_0$) | $n_{01}$=17 | $n_{00}$=686 |
| $P(h_1|E)$ | $n_{11}/(n_{01}+n_{11})$=0.830 | $n_{10}/(n_{00}+n_{10})$=0.0767 |
| $T(A_1^b|E)$ | 1 | $b'^*$=0.0924 |

According to Eq. (25) and Table 4, there is

$$b'^* = \frac{n_{10}}{n_{00} + n_{10}} \Big/ \frac{n_{11}}{n_{01} + n_{11}} = \frac{n_{10}(n_{01} + n_{11})}{n_{11}(n_{00} + n_{10})} \quad (27)$$

Hence, $b'^*$=0.0924, $b^*$=1-$b'^*$=1-0.0924=0.908, and $I(h_1{}^b, E)$ =0.923 bit. This information is equal to the KL information.

Consider inference $h_1$= "If a person has high triglyceride, then he also has fatty liver". Data from health examines of 142 people[3] are shown in Table 5.

**Table 5** The optimization of the degrees of disbelieve $b'$ about fatty liver

| numbers | fatty liver ($e_1$) | non-fatty liver ($e_0$) |
|---|---|---|
| high triglyceride ($h_1$) | $n_{11}$=25 | $n_{10}$=16 |
| low triglyceride ($h_0$) | $n_{01}$=41 | $n_{00}$=60 |
| $T(A_1|E)$ | 1 | $b'^*$=0.556 |

According to Table 5, the DOC $b^*$=0.444 and the information $I(E; h_1{}^{b^*})$ =0.025 bit. Comparing the two inferences, obviously the inference about birds is more informative and more believable then the inference about fatty liver.

## 6. Raven Paradox and Rapid HIV Tests

Hempel (1945) describes the paradox in terms of the hypothesis:

(1) "All ravens are black".

It is equivalent to:

(2) "All non-black things are not ravens".

A non-black non-raven thing ($e_{00}$) such as a white chalk supports (2) and hence also supports (1). Yet, according common knowledge, a white chalk is irrelative to (1). So, there is a paradox. There have been many articles about this paradox. (Good, 1960; Scheffler and Goodman, 1972; Maher, 1999; Fitelson and Hawthorne, 2010).

Now let $s_1$= "$E$ is in $S_1$", $s_2$ = "$E$ is in $S_2$", and $s_1$->$s_2$ = "If $E$ is in $S_1$ then $E$ is in $S_2$", and so on. To resolve the paradox, there are two ways generally. One is to deny **Equivalence Condition** (**EC**) ($s_1$->$s_2$ is equivalent to ¬$s_2$->¬$s_1$); another is to deny **Irrelevance** ($e_{00}$ is irrelative to $s_1$->$s_2$). Most researchers like Hempel affirm the **EC** and deny the **Irrelevance**, and believe $e_{11}$ can support $h_1$ better than $e_{00}$ (Fitelson and Hawthorne, 2010). Like Scheffler and Goodman (1972), the author also denies the **EC** and emphasizes falsification. Unlike almost all researchers, the author argues that for inference $h_1$=$s_1$->$s_2$, in many cases, an evidence $e_{00}$ can support $h_1$ better than $e_{11}$.

---

[3] From figue 1 in this page: http://www.ncbi.nlm.nih.gov/pmc/articles/PMC2829436/. Fatty liver patients include borderline and NASH groups.

For a general universal hypothesis, there are four kinds of inferences as shown in Table 6. Each inference has a particular pair of positive example and counterexample. For $h_1 = s_1$->$s_2$, the positive example is $e_{11}$ and the counterexample is $e_{10}$. For $h_2 = \neg s_2$->$\neg s_1$, the counterexample is also $e_{10}$, but the positive example is $e_{00}$. In Mathematical Logic, $h_1 = h_2$; however, in fuzzy logic or probability logic, $h_1^{b1}$ and $h_2^{b2}$ should be different because their proportion of counterexamples to positive examples are different.

**Table 6** Four kinds of inferences and their positive examples and counterexamples

|  | $h_3 = s_2$->$s_1$ | $h_2 = \neg s_2$->$\neg s_1$ |
|---|---|---|
| $h_1 = s_1$->$s_2$ | positive example $e_{11}$ | counterexample $e_{10}$ |
| $h_0 = \neg s_1$->$\neg s_2$ | counterexample $e_{01}$ | positive example $e_{00}$ |

According Eq. (27), $h_1$ and $h_2$ should be given different degrees of disconfirmation (the optimized DOD):

$$b_1^{'*} = \frac{n_{10}}{n_{00} + n_{10}} \Bigg/ \frac{n_{11}}{n_{01} + n_{11}} \quad (28)$$

$$b_2^{'*} = \frac{n_{10}}{n_{11} + n_{10}} \Bigg/ \frac{n_{00}}{n_{01} + n_{00}} \quad (29)$$

The partial derivatives of $b_1* = 1 - b_1'*$ with respect to $n_{11}$ and $n_{00}$ respectively are:

$$\frac{\partial b_1^*}{\partial n_{11}} = \frac{n_{10} n_{01}}{(n_{00} + n_{10}) n_{11}^2} \quad (30)$$

$$\frac{\partial b_1^*}{\partial n_{00}} = \frac{n_{10}(n_{01} + n_{11})}{n_{11}(n_{00} + n_{10})^2} \quad (31)$$

The two partial derivatives tell us how much the two different positive examples $e_{11}$ and $e_{00}$ raise respectively the DOC of $h_1$. Assume $n = n_{11} + n_{00} + n_{10} + n_{01}$ is much greater than 1. Then $\partial b_1^* \big/ \partial n_{11}$ is the increment of $b_1*$ raised by a new single evidence $e_{11}$, and $\partial b_1^* \big/ \partial n_{00}$ is the increment of $b_1*$ raised by new $e_{00}$.

Why do people think that a non-black non-raven thing (such as a white chalk) is irrelative to "All ravens are black"? Firstly, people have never seen non-black raven, which means $n_{10} = 0$ and $n_{11} > 0$. So $b_1* = 1$, no matter how big $n_{00}$ is. If $E \in A$ that only includes birds and $h_1 =$"Most swans are white", then it should be acceptable by most researchers that an evidence $e_{00}$ (a non-white non-swan thing) supports $h_1$. If $h_1$ refers to the result of rapid HIV + (positive) and means "This person has infected by HIV probably", an evidence $e_{00}$ (a noninfected person whose rapid HIV test shows negative) should also supports $h_1$.

Secondly, as some researchers have pointed out (Good, 1960), for "All ravens are black", the number $n_{00}$ of $e_{00}$ (non-black non-raven thing) is very big so that an evidence $e_{00}$ can hardly affect $h_1$ even if there are some counterexamples. Eq. (31) supports this idea since when $n_{00}$ is much bigger then $n_{01}$, $\partial b_1^{*}/\partial n_{00}$ is close to 0.

Almost all researchers believe that evidence $e_{11}$ can support $h_1$ better than $e_{00}$ (Fitelson and Hawthorne, 2010). However, according to Eq. (30) and (31), we can draw a different conclusion. When $(1+n_{11}/n_{01})n_{11}-n_{10}>n_{00}$, $\partial b_1^{*}/\partial n_{00} > \partial b_1^{*}/\partial n_{11}$. For example, assuming $n_{11}=n_{01}=10n_{10}$, then when $n_{11}>n_{00}/1.9$, $\partial b_1^{*}/\partial n_{00} > \partial b_1^{*}/\partial n_{11}$. This means that in many cases, $e_{00}$ can support $h_1$ better than $e_{11}$. This conclusion is unexpected.

We may also use Eq. (28) to get similar conclusion. Assuming $n_{00}$ is much greater than $n_{10}$ and $n_{11}=n_{01}$, when $n_{00}$ doubles, $b_1$'* reduces almost half; yet, when $n_{11}$ doubles, $b_1$'* only reduces 1/4. So, in many cases, the increment of $n_{00}$ can reduce the degree of disconfirmation of $h_1$ faster than the increment of $n_{11}$.

Let's use OREQuick HIV tests[4] to show the importance of $e_{00}$ as the evidence for confirmation of $h_1$. The result of a HIV Test is either + (positive) or – (negative). The $h_1$ becomes + and $h_0$ becomes -. For a given patient with HIV $e_1$, the conditional probability of + is $P(+|e_1)$, which is called sensitivity by medical industry. For HIV-noninfected people, the conditional probability of - is $P(-|e_0)$, which is called specificity (as shown in Table 7).

**Table 7** $P(+|E)$ and $P(-|E)$ for OREQuick HIV Tests

|  | with HIV ($e_1$) | without HIV ($e_0$) |
|---|---|---|
| $P(+|E)$ | sensitivity=0.917 | 1-specificity=0.001 |
| $P(-|E)$ | 1-sensitivity=0.083 | specificity=0.999 |

According to Eq. (24), $b_1$'*=$P(+|e_0)/P(+|e_1)$ = (1-specificity)/sensitivity =0.001/0.917≈0.0011; $b_1$*=1-0.0011=0.9989; $I(E; +^{b1*})$ =5.52 bits. For testing result -, $b_0$'*= (1-sensitivity)/specificity=0.083/0.999=0.083; $b_0$*=1-0.083=0.917; $I(E; -^{b0*})$ =0.04 bit.

It is easy to notice that specificity is more important to raise the DOC of $h_1$=+ than sensitivity. For example, even if sensitivity is 0.5 (or 0.1), as long as specificity is 1, then $b_1$* will be 1, which means that it is absolutely believable to diagnose AIDS according to +. If specificity is 0.5, even if sensitivity is 1, then $b_1$* will be only 0.5, which means that it is half believable to diagnose AIDS according to +. Similarly, we can prove that sensitivity is more important to raise the DOC of - than specificity. The reason is that less counterexamples are more important than more positive examples for us to believe a hypothesis.

---

[4] http://www.oraquick.com/taking-the-test/understanding-your-results

When $n$ is big enough, specificity=$n_{00}/(n_{00}+n_{10})$ and sensitivity=$n_{11}/(n_{11}+n_{01})$. So, $n_{00}$ is related to specificity. That is why $e_{00}$ can support $h_1$ better than $e_{11}$ in many cases.

Likelihood Ratio (LR) (=sensibility/(1-specificity=$1/b'$*) is used by medical community to tell how good the test-positive is. It is easy to find that $b*$ is positively related to LR. The difference is the upper limit of $b*$ is 1 so that $b*$ is suitable as DOC. In addition, $b'$* can be used to predict the probability in which the testee has disease.

Statisticians and some doctors often argue about the reliability of medical tests. If the prior probability $P(e_1)$ or $P$(HIV) of a person with HIV is about 0.004, then according to Bayes' formula, the posterior probability $P(e_1|+)$=0.917*0.004/(0.001*0.996+0.917*0.004)=0.786, which is not high enough. If the prior probability $P(e_1)$ is 0.0001 instead of 0.004, the posterior probability $P(e_1|+)$ will be 0.08. Can we still believe the test? The author answers "yes" as most doctors. Actually this degree $b_1*$=0.9989 is irrelative to the prior probability $P(e_1)$ or which group of people the testee belongs to. To predict the probability in which the testee has AIDS by +, most statisticians are right; yet, to believe the HIV Test or not, most doctors are right. However, to predict the probability, we may use the semantic bayes' formula Eq. (26). For example, for a testee who belongs to high-risk group of people, $P(e_1)$ =0.1. Then $P(e_1|h_1^{b*})$ =0.1/(0.1+0.0011*0.9) =0.991. This result is the same as $P(e_1|+)$ obtained by Bayes' formula. Yet, Eq. (26) is simpler.

The popular confirmation measures, such as $d(H, E)$ (Earman, 1992) and $s(H, E)$ (Christensen, 1999; Joyce, 1999), use LP and conditional LP. One can get the LP from prior knowledge and the conditional LP from one or two evidences, without using sensibility and specificity. Yet it is difficult to apply them to medical tests. One may argues that they are used for the increments of DOC. Yet, as increments, they are too big. If more evidences come, how do we deal with the measures? As comparison, this paper uses Eq. (15) and (16) for the increments of DOC.

If we replace LP with SP in $d(H, E)$, $d(+, E)$ will decreases with $P(e_1)$ increasing. This is unreasonable. When specificity is 1 and sensibility is 0.1, to predict the probability $P(e_1|+)$ by Bayes' formula or the semantic Bayes' formula, $P(e_1|+)$=1, which means the prediction is 100% correct. Yet, both $d(+, E)$ and $s(+, E)$ are less than 0.1. For 100% accuracy predicted by +, so low DOCs are unreasonable.

Why no others have proposed the $b*$? The reason might be that according to Eq. (23), $b*$ might be -∞, and its upper limit (1) and lower limit (-∞) are asymmetrical. Yet, using the semantic information method, the negative DOC needs another formula and its lower limit is -1.

# 7. Negative DOC with "All swans are white" as Example

First, consider the positive DOC for $h_1$="All swans are white". Assuming that the posterior ADOD $Q_0/Q_1$=0.01/0.99 is less than the prior ADOD $P_0/P_1$=0.2/0.8 (see Table 8), then the DOC $b*$ is positive. According to Eq. (24), we get

$$b'*= (Q_0/Q_1)/(P_0/P_1) = (0.01/0.99) =0.0404; b*=1-b'*=0.9596.$$

**Table 8** Positive DOC for "All swans are white" ($b^*>0$)

|  | white swan ($e_1$) | non-white swan ($e_0$) | average information |
|---|---|---|---|
| $P(E)$ | 0.8 | 0.2 | |
| $P(E|h_1)$ | 0.99 | 0.01 | $I_{KL}(E; h_1)$=0.2611 bit |
| $T(h_1|E)$ | 1 | 0 | $I(E; h_1)$=-∞ |
| $T(h_1{}^{b^*}|E)$ | 1 | $b$'*=0.0404 | $I(E; h_1{}^{b^*})$=0.2611 bit |

However, for a lie or wrong hypothesis, such as "All swans are not white", or a prediction from a stock commentator who is seen as a contrary indicator, or a hypothesis with excessive affirmation, we may modify the DOB into negative value to get more average semantic information.

When a negative DOB $b$ ($b<0$) is given to a general hypothesis $h_j$, $h_j$ becomes $h_j{}^b$, whose the truth function is defined as

$$T(h_j{}^b |E) = 1 + bT(h_j|E), \text{ for } b<0 \qquad (32)$$

Now the DOD $b$'=1-$|b|$=1+$b$.

Figure 5 illustrates how positive $b$ and negative $b$ affect $T(h_1{}^b |E)$.

Now consider that counterexamples increase so that the DOC $b^*$ of $h_1$="All swans are white" is negative. Assuming the prior ADOD is 0.01/0.99, the posterior ADOD increases to 0.05/0.95 after more black swans occur (as shown in Table 10).

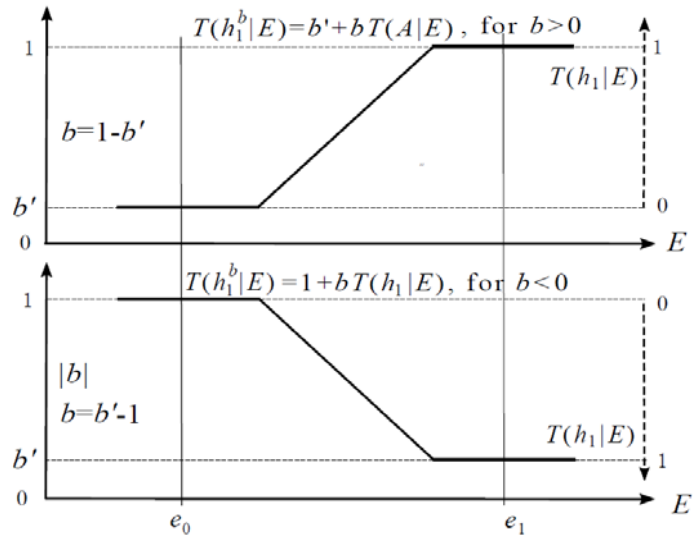

**Figure 8** How positive $b$ and negative $b$ affect $T(h_1{}^b |E)$

**Table 9** Negative DOC for "All swans are white" ($b^*<0$)

| | white swan ($e_1$) | non-white swan ($e_0$) | average information |
|---|---|---|---|
| $P(E)$ | 0.99 | 0.01 | |
| $P(E|h_1)$ | 0.95 | 0.05 | $I_{KL}(E; h_1)$= 0.060 bit |
| $T(h_1|E)$ | 1 | 0 | $I(E; h_1)$=-$\infty$ |
| $T(h_1{}^{b^*}|E)$ | $b'^*$= 0.192 | 1 | $I(E; h_1{}^{b^*})$=0.060 bit |

According to Eq. (3) and (32), we get $T(A_1) = P_0 + b'P_1$. According to Eq. (9) and (32), we get

$$I(E; h_1{}^b) = Q_0 \log \frac{1}{P_0 + b'P_1} + Q_1 \log \frac{b'}{P_0 + b'P_1} \qquad (33)$$

To optimize $b$, $P_0>0$ is needed. Assuming $P_0>0$ and derivative d$I(E; h_1{}^b)$/d$b'$=0, then

$$b'^* = (P_0/P_1)/(Q_0/Q_1) \qquad (34)$$

where $(P_0/P_1)/(Q_0/Q_1)$ must be less than 1, otherwise Eq. (22) is needed. According to Eq. (32), we have

$$b^* = b'^*-1 = (P_0/P_1)/(Q_0/Q_1)-1 \qquad (35)$$

Therefore, when the posterior ADOD increases from 0.01/0.99 to 0.05/0.95, $b'^*$= (0.01/0.99)/(0.05/0.95) =0.192; $b^*$= $b'^*$-1=-0.808.

According to Eq. (35), if $Q_0>Q_1$, the posterior ADOD should be negative and equal to $Q_1/Q_0$-1. Similarly, if $P_0>P_1$, the prior ADOD is equal to $P_1/P_0$-1<0.

It is not that only Eq. (34) is used to get negative DOC. The Eq. (22) can also be used to get negative DOC for excessive negation. The Eq. (34) can also be used to get positive DOC for proper negation (as shown in Table 10).

**Theorem 1**. A denial hypothesis $h_0(E) = \neg h_1(E)$ with negative DOB $b_0$ is equal to the affirmative hypothesis $h_1(E)$ with positive DOB $|b_0|$, i. e.,

$$T(h_0{}^{b0}|E) = T(h_1{}^{|b0|}|E), \text{ for } Q_0/Q_1 \leq P_0/P_1 \text{ and } b_0<0 \qquad (36)$$

**Proof**: According to fuzzy logic (Zadeh, 1965), $T(h_0|E) =1-T(h_1|E)$. According to Eq. (20) and the fuzzy logic, $T(h_0{}^{b0}|E) =1+b_0T(h_0|E) =1+b_0(1-T(h_1|E)) =1+b_0-b_0T(h_1|E)$. According to Eq. (20), $1+b_0-b_0T(h_1|E) =1-|b_0|+|b_0|T(h_1|E) = T(h_1{}^{|b0|}|E)$. Hence $T(h_0{}^{b0}|E) = T(h_1{}^{|b0|}|E)$.

**Table 10** DOC in 4 cases and the meanings of hypotheses with DOC

| Initial hypotheses | The ratio of counterexamples decreases: $Q_0/Q_1 \leq P_0/P_1$ | The ratio of counterexamples increases: $Q_0/Q_1 > P_0/P_1$ |
|---|---|---|
| | $b'* = (Q_0/Q_1)/(P_0/P_1)$ | $b'* = (P_0/P_1)/(Q_0/Q_1)$ |
| $h_1(E)$="All swans are white" | $b* = 1-(Q_0/Q_1)/(P_0/P_1)>0$ for proper affirmation, $h_1{}^{b*} \approx$ "There are more white swans than we expect" | $b* = (P_0/P_1)/(Q_0/Q_1)-1<0$ for excessive affirmation, $h_1{}^{b*} \approx$ "There are less white swans than we expect" |
| $h_0(E)$="All swans are not white" | $b_0* = (Q_0/Q_1)/(P_0/P_1)-1=-b*<0$ for excessive negation, $h_0{}^{b0*} \approx$ "There are less non-white swans than we expect"="There are more white swans than we expect"$\approx h_1{}^{|b0*|}=h_1{}^{b*}$ | $b_0* = 1-(P_0/P_1)/(Q_0/Q_1)=|b*|>0$ for proper negation, $h_0{}^{b0*} \approx$ "There are more non-white swans than we expect"="There are less white swans than we expect"$\approx h_1{}^{-b0*}=h_1{}^{b*}$ |

**Theorem 2**. A denial hypothesis $h_0(E)=\neg h_1(E)$ with positive DOB $b_0$ is equal to the affirmative hypothesis $h_1(E)$ with negative DOB $-b_0$, i. e.

$$T(h_0{}^{b0}|E) = T(h_1{}^{-b0}|E), \text{ for } Q_0/Q_1>P_0/P \text{ and } b_0>0 \quad (37)$$

**Proof**: $T(h_0{}^{b0}|E)= 1-b_0+b_0T(h_0|E) =1-b_0+b_0(1-T(h_1|E))$ $=1-b_0T(h_1|E)=1+(-b_0)T(h_1|E)=T(h_1{}^{-b0}|E)$.

The adjustment of DOB to increase its average information cannot be applied to all hypotheses or predictions. Most blind guesses do not convey positive average information, no matter how their DOB are adjusted. For example, if someone always predicts the rises or falls of stock markets by throwing a coin, then the DOC of his prediction can only be 0. Most wrong hypotheses still cannot convey meaningful positive information even if their DOBs are adjusted into negative values. For example, wrong prediction "Tomorrow is the end of the world" with any DOB can only convey information close to or less than 0.

## 8. The Degree of Confirmation of General Hypotheses

In Section 5, a fuzzy universal hypothesis is be formalized as $h_1{}^b=$"For all $E$, if $E \in S_1$, then $E \in A_1$". Now consider weather forecasts, GPS, and various fuzzy hypotheses, including fuzzy inferences and predictions. A general fuzzy hypothesis is formalized as:

$$h_j \text{ or } h_j{}^b = \text{"For all } E \text{ and } Z, \text{ if } Z \in C_j, \text{ then } E \in A_j\text{"}$$

GPS is a good example of general hypotheses. Now $E$ denotes the real position of a GPS device, $H$ denotes the position pointed by GPS arrow. If $E=e_i$ and $H=h_j=\hat{e}_j$, then $\hat{e}_j$ is the estimation of $e_i$ by GPS according to condition $Z$ (the distances to three or more satellites and other factors).

The initial hypothesis $h_j$ may be fuzzy or non-fuzzy. By making some assumptions for simplicity so that the initial hypothesis is non-fuzzy, we can also calculate the DOC $b^*$ of $\hat{e}_j$ provided by GPS as above.

Circular Error Probability (CEP)[5] is often used to express the accuracy of GPS. The CEP=10 meters means that for a given position $e_i$ of GPS device, the probability in which the estimation $h_j = \hat{e}_j$ from $e_i$ is not farther than 10 meters is 0.5. Let $S_i$ denote the circle with 10 meter radius surrounding center $e_i$. Then CEP=10 meters means that for give $e_i$, the probability of correct estimations ($\hat{e}_j$ in $S_i$) is $\sum_{\hat{e}_j \in S_i} P(\hat{e}_j \mid e_i) = 0.5$, the probability of wrong estimations is 1-0.5=0.5. Let's simply assume that $P(\hat{e}_j \mid e_i)$ for all correct estimations are the same and equal to $p_1$, and $P(\hat{e}_j \mid e_i)$ for all wrong estimations are the same and equal to $p_0$; the number of all positions in a circle with 10 meter radius is $n$, and the number of all possible positions is $1000n$. Then there are $p_1 = 0.5/n$, and $p_0 = 0.5/(999n)$. Hence, according to Eq. (24) and (25), $b'^* = p_0/p_1 = 0.5/(999n)/(0.5/n) = 1/999$; $b^* = 1 - b'^* = 998/999$.

If the initial hypothesis $h_j$ is fuzzy, we may seek the DOB $b^*$ that makes the average semantic information reach the maximum:

$$b^* = \arg\max_b I(E; h_j^b) = \arg\max_b \sum_i P(e_i \mid h_j) \log \frac{T(h_j^b \mid e_i)}{T(h_j^b)} \quad (38)$$

which is the definition of the DOC of a general hypothesis.

Another popular measure for GPS accuracy is Distance Root Mean Square (DRMS)[6]. DRMS=10 means that the standard deviation between $\hat{E}$ and $e_i$ is 10 meters. Generally we consider that $P(\hat{E} \mid e_i)$ as a normal distribution:

$$P(\hat{E} \mid e_i) = k \exp[-|\hat{E} - e_i|^2 / (2d^2)] \quad (39)$$

where $k$ is normalized coefficient; $e_i$ is the real position of the GPS device; $\hat{E}$ is the estimation of $e_i$; $d$ is DRMS or standard deviation which implies the precision of GPS. The DRMS=10 means that possibility in which the deviation within 10 meters is 65%. In this case, the initial hypothesis $h_j$ is fuzzy. Assume Eq. (39) is tenable for different $E$, hence there is

$$P(\hat{E} \mid E) = k \exp[-|\hat{E} - E|^2 / (2d^2)] \quad (40)$$

However, the above distribution is only for ideal estimation. The actual estimation may be the function of condition $Z$, i. e., $\hat{E} = f(Z)$. The deviation distribution may be

$$P(\hat{E} \mid E) = k \exp[-|\hat{E} - \Delta e - E|^2 / (2d^2)] + c \quad (41)$$

---

where $\Delta e$ is systematic deviation ($\Delta e$=0 means the highest accuracy); $c$ mean that there are more long distance deviations, which come from wrong map or systematic failure. It is the $c$ that determines the DOC $b^*$ of $\hat{E}$. To confirm $H=\hat{E}$, the average semantic information becomes semantic mutual information:

$$I(E;H^b) = \sum_j P(\hat{e}_j) \sum_i P(e_i \mid \hat{e}_j) \log \frac{T(h_j^b \mid e_i)}{T(h_j^b)} \quad (42)$$

Assuming the optimized truth function is

$$T(\hat{e}_k^{b^*} \mid E) = b^* \exp[-|\hat{e}_k - E|^2] / (2d^{*2}) + 1 - b^*, \ k=1, 2 \ldots (43)$$

by using Eq. (34) and (18), we can derive that when $\hat{e}_k = \hat{e}_j - \Delta e$ ($j$=1, 2…), $d^*=d$, $b^*$=1-$c/(k+c)$, the semantic mutual information $I(E; H^{b^*})$ reaches the maximum. From this example, it is easy to find that we need to consider three factors: accuracy, precision, and DOC for selecting a hypothesis from many with information criterion. If $E$ is not equally probable, LP $T(A_j)$ is a better measure than $d$ as precision because $T(A_j)$ is also related to $P(E)$ and determines the testing severity. Actually, optimizing $T(H|E)$ with $P(H|E)$ is optimizing semantic channel so that it matches Shannon's channel.

If a GPS user wants to confirm or optimize the estimation $\hat{e}_j$ (or $\hat{E}$), he needs $P(\hat{e}_j|E)$, which cannot be obtained directly. Since when $E$ is equally probable, i. e., $P(E)=a$ ($a$ is a constant), we could derive $P(\hat{E})=a$, and $P(\hat{e}_j|E)=P(E|\hat{e}_j)=P(E, \hat{e}_j)/a$. So, the user may put the GPS device at different positions in equal probability to record $E$ and $\hat{E}$. Then he could get $P(\hat{E}, E)$ and $P(\hat{e}_j|E)$, $j$=1 2… $n$. If the probability distribution $P(E|\hat{e}_j)$ of samples is obtained when $P(E)\neq$constant, one could, according to Eq. (19), use $P(E|\hat{e}_j)/P(E)$ to replace $P(\hat{e}_j|E)$ to confirm $\hat{e}_j$.

In comparison with MLE, MSIE has two advantages: 1) The MSIE can be used in cases where source $P(E)$ is variable; 2) Generally, $P(h_j|E)$ is more regular than $P(E|h_j)$ so that it is easier to construct $T(A_j|E)$ proportional to $P(h_j|E)$ than to construct $P(E|A_j)$ close to $P(E|h_j)$. For example, $P(E|A_j)$ predicted by GPS will not be a normal distribution and changes from place to place because roads are irregular; yet, $P(h_j|E)$ is an approximately normal distribution and almost changeless.

## 9. Conclusions and Discussions

In this paper, the semantic information measure is used as the main tool for falsification and confirmation. This measure is compatible with Shannon's theory, Popper's theory, and Fisher's likelihood method. With generalized Kullback-Lribler formula, we can use objective sampling distribution to test subjective probability prediction (likelihood function); use objective selecting rule function (or Shannon channel) to confirm subjective truth function (or semantic channel). The basic conclusions about falsification and confirmation are: 1) For the falsification (including test, selection, and optimization) of hypotheses, we need semantic information as criterion, which means that the more prior precise and unexpected and the more posterior accurate and believable a hypothesis is, the more information it conveys and hence the better it is. 2) Confirmation, for which less counterexamples are more important than more positive examples, is to get the optimized degrees of belief in hypotheses to increase average semantic information, and hence is a helper of falsification.

MSIE and MLE use the same semantic information criterion in essence and compatible with Shannon's information theory. But, the MSIE is more suitable to cases where source $P(E)$ is variable, channel $P(H|E)$ is stable and regular, and the amount of ample are huge.

The MSIE requires that samples are independent and their distribution is stable. If there are only fewer samples, the distribution $P(E|h_j)$ may be unstable and will result in over fitting. To resolve this problem, we may decrease the degree of confirmation obtained according to the samples. Can we set up the relationship between the number of samples and the degree of confirmation, or prior limit the extents of precision and degree of belief in cases with fewer samples? This is a question that needs further study.

The Maximum A Posterior (MAP) estimation (DeGroot, 1970) may get better results when there are fewer samples. Besides likelihood, MAP also use prior probability $P(\theta)$ or $P(h_j)$. Yet MAP is not compatible with Shannon's information theory. Is the prior probability $P(\theta)$ statistical or logical? This could be a question. Similarly, the MSIE uses the prior probability distribution $P(E)$ and similar Bayes formula ($P(h_j|E) = \ldots$). In cases with fewer samples, how could we make full use of prior knowledge better？ This deserves further studies as well.

## Appendix: List of Symbols

$A=\{e_1, e_2\ldots\}$ is a set of evidences, $e_i$—one element in $A$

$E \in A$ is a variable

$A_j$ is a fuzzy subset of $A$; the elements *in* $A_j$ make hypothesis $h_j$ be true

$B= \{h_1, h_2\ldots\}$ is a set of hypotheses; $h_j$ is one element in $B$; $H \in B$ is a variable

$C= \{z1, z_2\ldots\}$ is a set of conditions; $C_j$ is a subset of $C$. When $Z \in C_j$, $h_j$ is selected.

$h_j(e_i)$—a proposition, such as $h_j(e_i) =$"$e_i \approx e_j$"; $h_j(E)$—a predicate, such as, $h_j(E) =$"$E \approx e_j$"

$h_j^{\ b}$—$h_j$ whose degree of belief is $b$

$T$—logical probability or true value; $T(h_j) = T(A_j)$—logical probability of hj= $h_j(E)$

$T(h_j|e_i) = T(A_j|e_i)$ — true value of a proposition $h_j(e_i)$

$T(h_j|E) = T(A_j|E)$—truth function of a predicate $h_j(E)$

$P$- Statistical probability; $P(E)$—prior probability (function), or source, $P(e_i)$—the prior probability of $e_i$

$P(h_j)$—statistical or selective probability of $h_j$

$T(h_j) = T(A_j)$—logical probability or average true value of $h_j(E)$

$P(E|A_j) = P(E)T(A_j|E)/T(A_j)$—9 probability, or theoretical prediction

$P(E|h_j) = P(E)P(h_j|E)/P(h_j)$—sampling distribution, inverse condition probability

$P(h_j|E)$—conditional probability, selective probability (function) of $h_j$

$e(t)$—the evidence under t-th condition $z(t)$, $t=1, 2\ldots w$. $e(t) \in A$

Assume the number of $e_i$ in $\{e(t), e(t)\ldots e(w)\}$ is $w_i$ and $w$ is enough big, then $P(e_i|C_j) = w_i/w$.

$P(E|C_j) = P(E|Z \in C_j) = P(E|h_j)$ is the sampling distribution on $A$ under conditions in $C_j$.

$P(E|A_j)$—likelihood

$b$— Degree of Belief (DOB)；$b'=1-|b|$--Degree of Disbelief (DOD)

$b^*$-- Degree of Confirmation (DOC), i. e., optimized DOB

$b'^*$--optimized degree of disbelief

ADOD—Absolute degree of disbelief

$I(e_i; hj)$—semantic information conveyed by $h_j$ about $e_i$

$I(E; h_j)$—average information conveyed by $h_j$ about $E$

KL--Kullback-Leibler

GKLF—generalized Kullback-Leibler Formula

LP—logical probability;

SP—statistical probability or selected probability

SIM—Semantic Information Measure

MLE—Maximum likelihood estimation

MSIE—Maximum semantic information estimation